\documentclass[letterpaper,twocolumn,10pt]{article}
\usepackage{usenix2022_SOUPS}

\usepackage{microtype}
\usepackage[symbol]{footmisc}

\usepackage{tikz}
\usepackage{amsmath}
\usepackage{amssymb}
\usepackage{amsthm}
\usepackage{mdframed}
\usepackage{comment}
 \newtheorem{thm}{Theorem}[section]

 \theoremstyle{definition}
 \newtheorem{defn}[thm]{Definition}

\newcommand{\be}{\begin{enumerate}}
\newcommand{\ee}{\end{enumerate}}

\usepackage{breakurl}
\begin{document}
%-------------------------------------------------------------------------------

%don't want date printed
\date{}

\title{Fair Context-Aware Privacy Threat Modelling}

% if you leave this blank it will default to a possibly ugly attempt 
% to make the contents of the \author command below into a string
\def\plainauthor{Author name(s) for PDF metadata. Don't forget to anonymize for submission!}

%for single author (just remove % characters)
\author{
{\rm Saswat Das\footnote[1]{Eq}}\\
National Institute of Science Education and Research,\\ 
An OCC of Homi Bhabha National Institute, India
\and
{\rm Rakshit Naidu\footnote[1]{Eq}}\\
Carnegie Mellon University
% copy the following lines to add more authors
% \and
% {\rm Name}\\
%Name Institution
} % end author

\maketitle
\thecopyright

% \let\thefootnote\relax\footnotetext[1]{*Equal Contribution}

%-------------------------------------------------------------------------------
\begin{abstract}
%-------------------------------------------------------------------------------
Given the progressive nature of the world today, fairness is a very important social aspect in various areas, and it has long been studied with the advent of technology. %In this short paper, we introduce \lq\lq fair privacy threat modelling" which examines how modelling privacy threats need to account for factors with equal parity. 
To the best of our knowledge, methods of quantifying fairness errors and fairness in privacy threat models have been absent. To this end, in this short paper, we examine notions of fairness in privacy threat modelling due to different causes of privacy threats within a particular situation/context and that across contexts.
\end{abstract}

%-------------------------------------------------------------------------------
\section{Introduction}
%-------------------------------------------------------------------------------
\footnotetext[1]{Equal Contribution}

Fairness and Privacy Threat Modelling (PTM) have long been studied in isolation. The reader is referred to these works~\cite{Jacobs2021MeasurementAF, Friedler2021TheO} and~\cite{Gholami2014PrivacyTM, Hussain2014THREATMM} respectively. 

% In this paper, we provide definitions for fair, context-aware privacy threat modelling. 
Privacy threat modelling involves thinking of mitigation strategies and solutions against privacy threats in various contexts.
% In this work, we specifically define our problem for a single privacy threat (which in the real-world, would be the outcome) which can be modelled across different contexts. 
For example, let us consider the privacy threat as the loss of the secret password of the victim. This incident would have different implications if the victim's password was lost within an office setting than that in a park/caf\'e with public WiFi. That is, it is more obvious for the victim to lose his password in public scenarios due to shoulder surfing than in private, work or office settings. Hence, we believe it is also important to fairly consider contexts in which the privacy threat occurs and assign probabilities to each cause that affects the threat fairly, based on the context in which the threat occurs. In this work, we formally define notions of fairness for context-aware privacy threat modelling. 

% \textit{Comment: Add citations, Cronk's work perhaps, some citations about possible unfairness in threat modelling perhaps or how unfairness may arise or even impact the privacy of the system. A mini literature review, as it were.}

We consider privacy threats in various situations/contexts %($\{X_1, X_2, \cdots, X_m\} \in \mathcal{C}$) 
given a set of possible causes% ($\{c_1, c_2, \cdots, c_n\} \in \chi$)
, based upon which the outcomes may vary\footnote[2]{Hence, we consider a set of contexts that is separate from the set of all outcomes; this allows us to study contexts based on different universes of possible causes. The reason for this will become apparent in section \ref{sec:def}.}. In a privacy threat model, the causes can be weighted in terms of how likely they are to lead to a given outcome. Given a set of causes from a universe of possible causes and a context, % chosen from $\chi$ and $\mathcal{C}$ respectively, 
we have a particular outcome, which is the real-word privacy threat we are modelling for.

It is possible that in a particular context, we may have a particular cause being given more weightage than it should be, owing to either suboptimal formulation of the context's conditions with respect to the set of all possible causes or the contribution of certain causes to the privacy risk may have been diminished or exaggerated~\cite{Abel2020}. Therefore, to perform fair privacy threat modelling, methods to quantify fairness errors owing to these factors would be beneficial, and we provide formal definitions in section \ref{sec:def}.%\emph{SD: Mannn you are the causality guy, check this and fix it as necessary :P}
%-------------------------------------------------------------------------------

\begin{figure}[ht]
    \centering
        
     \includegraphics[width=0.3\textwidth]{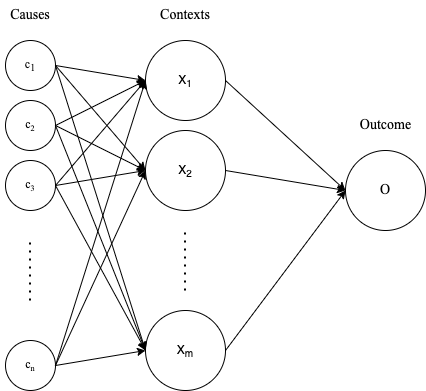}
     \caption{The set of all causes ($\{c_1, c_2, \cdots, c_n\} \subseteq \chi$) and the set of contexts ($\{X_1, X_2, \cdots, X_m\} \subseteq \mathcal{C}$) for a particular privacy threat outcome ($\mathcal{O}$).}
     \label{fig:fig1}
    % \vspace{-2ex}
\end{figure}

\section{Formal Definitions}
\label{sec:def}
%-------------------------------------------------------------------------------
%\subsection{Definitions}
Suppose we want to study a privacy threat model in a context $X$ from a set of contexts $\mathcal{C}$ using causes from a universe of all possible causes $\chi$ that affect that particular privacy threat. Then a threat model within a particular context $X$ can be seen as an assignment of probabilities to different causes in $\chi$ according to their contribution to the resulting outcome $\mathcal{O}%_{(X,\chi)}
$ within that context.

More precisely, given $\chi$ and $X$, we can assign a probability $p^\chi_c:=\Pr[\mathcal{O}%_{(X,\chi)}
|X,c,\chi]$ for every cause $c\in\chi$, yielding the probability vector $(p^\chi_c)_{c\in\chi}$. %Note that $\sum_{c\in\chi}p_c=1$. NB: Redundant, as I already stated that this is a probability vector.

Now unfairness can arise in these models with respect to each cause and across various contexts; we thus provide a way to mathematically quantify these fairness errors.
% \subsection{Unfairness due to Causes}
\begin{defn}{\emph{Fairness Error due to a Cause $c\in\chi$}}\\
The fairness error due to a cause $c\in\chi$, for an outcome $\mathcal{O}$ amd for every $X\in\mathcal{C}$, can be defined as
\[\mathcal{E}_{(\chi\setminus\{c\})}\overset{\Delta}{=}\sum_{X\in\mathcal{C}}\sum_{c'\in\chi\setminus\{c\}}\left|\Pr[\mathcal{O}%_{(X,\chi)}
|X,c',\chi]-\Pr[\mathcal{O}%_{(X,\chi\setminus\{c\})}
|X,c',\chi\setminus\{c\}]\right|.\]
\end{defn}

Here the fairness error for a cause $c\in\chi$ is given by the sum over all contexts $X\in\mathcal{C}$ of the $\ell_1$ norm distance between the probability vectors with the entry corresponding to $c$ omitted.
% \subsection{Unfairness due to a Context}
% In this we try to gauge, loosely speaking, the unfairness of a context. For this, we provide the following definition.
\begin{defn}{\emph{Fairness Error due to a Context $X\in\mathcal{C}$}}\\
For an outcome $\mathcal{O}$ and each context $X\in\mathcal{C}$, we define the fairness error for a context $X\in\mathcal{C}$ as
\[\mathcal{E}_X\overset{\Delta}{=}\sum_{c\in \chi}\sum_{c'\in \chi\setminus\{c\}}\left|\Pr[\mathcal{O}%_{(X,\chi)}
|X,c',\chi]-\Pr[\mathcal{O}%_{(X,\chi\setminus\{c\})}
|X,c',\chi\setminus\{c\}]\right|.\]
\end{defn}

That is, here the fairness error for the context $X$ is the sum over all $c\in\chi$ of $\ell_1$ norm distances between the probability vectors with only those entries \emph{not corresponding to $c$} of the probability vectors with respect to the set of causes being $\chi$ and $\chi\setminus\{c\}$ respectively.

\subsection{Notions of Fairness for Threat Models}
The aforementioned errors can serve as quantities to study and rank causes and contexts based on fairness, and can serve as a starting point for trying to mitigate substantial fairness errors with respect to a defined threshold, if any. In this vein, it would also help to define a way to quantify fairness error bounds.

\begin{defn}{\emph{$\lambda$-Causal Fairness}}\\
Given a universe $\chi$ of all possible causes, and a set $\mathcal{C}$ of all possible contexts within which privacy threats may occur, a privacy threat model is said to be $\lambda$-causally fair (for $\lambda\geq 0$) if \(\sup_{c\in\chi}\mathcal{E}_{\chi\setminus\{c\}}\leq\lambda.\)\footnote[3]{Here $\sup$ refers to the supremum/least upper bound of the set of values obtained in question.}
\end{defn}

\begin{defn}{\emph{$\gamma$-Contextual Fairness}}\\ 
Given a universe $\chi$ of all possible causes, and a set $\mathcal{C}$ of all possible contexts within which privacy threats may occur, a privacy threat model is said to be $\gamma$-contextually fair (for $\gamma\geq 0$) if
\(\sup_{X\in\mathcal{C}}\mathcal{E}_X\leq\gamma.\)
\end{defn}
%-------------------------------------------------------------------------------
\section{Discussion}
Based on the causal and contextual error bounds for a certain cause or context respectively, a diagnosis of the unfairness incurred can be done, and subsequent steps for mitigation of the same can be taken. If the causal fairness error bound is substantially high, then one can look at the fairness error due to various causes individually and try to make changes to the most unfair causes in $\chi$ or add/subtract causes from $\chi$ to reduce this error. Similarly, one can improve the contextual fairness error bound on $\mathcal{C}$.
%-------------------------------------------------------------------------------
%On what a high fairness bound might mean in both senses, and directions for mitigation maybe?
%-------------------------------------------------------------------------------
\section{Conclusion and Future Directions}
%-------------------------------------------------------------------------------
We have thus attempted to provide certain methods of quantifying the fairness errors present in privacy threat models and some notions of fairness in that regard. This is but the first step in studying the fairness of privacy threat modelling. Some further directions to extend this can include deducing further mathematical properties based on the provided definitions, or designing diagnosis and mitigation strategies based on these concepts/notions. This can overall make for better and more reasonable threat modelling across various contexts.  
FAIR \cite{Cronk2021FAIRIEEE} provides a way to quantify privacy risk as \textit{(threat\_frequency * harm\_magnitude)}. We believe our work can also influence the FAIR framework making threat modelling more context-aware by quantifying privacy risk as \textit{[(threat\_frequency * harm\_magnitude) | context]}. 

%-------------------------------------------------------------------------------
\bibliographystyle{plain}
\bibliography{main}

%%%%%%%%%%%%%%%%%%%%%%%%%%%%%%%%%%%%%%%%%%%%%%%%%%%%%%%%%%%%%%%%%%%%%%%%%%%%%%%%
\end{document}